\documentclass[aps,prb,twocolumn,superscriptaddress,floatfix]{revtex4}
\usepackage{psfig}
\usepackage{graphicx}
\usepackage{epsfig}
\bibstyle{apsrev.bib}

\begin{document}
\input epsf.sty

\title{Disorder-induced rounding of the phase transition in the large $q$-state Potts model}

\author{M. T. Mercaldo}
\affiliation{
Centre de Recherches sur les Tr\'es Basses
Temp\'eratures\thanks{U.P.R. 5001 du CNRS, Laboratoire conventionn\'e
avec l'Universit\'e Joseph Fourier}, B. P. 166, F-38042 Grenoble,
France}
\affiliation{Dipartimento di Fisica ``E.R. Caianiello''  and  Istituto
Nazionale per la Fisica della Materia, Universit\`a degli Studi di
Salerno, Baronissi, Salerno I-84081, Italy}
\author{J-Ch. Angl\`es d'Auriac}
\affiliation{
Centre de Recherches sur les Tr\'es Basses
Temp\'eratures\thanks{U.P.R. 5001 du CNRS, Laboratoire conventionn\'e
avec l'Universit\'e Joseph Fourier}, B. P. 166, F-38042 Grenoble,
France}
\author{F. Igl\'oi}
\affiliation{
Research Institute for Solid State Physics and Optics,
H-1525 Budapest, P.O.Box 49, Hungary}
\affiliation{
Institute of Theoretical Physics,
Szeged University, H-6720 Szeged, Hungary}

\date{\today}

\begin{abstract}

The phase transition in the $q$-state Potts model with homogeneous ferromagnetic
couplings is strongly first order for large $q$, while is rounded in the presence
of quenched disorder. Here we study this phenomenon on different two-dimensional
lattices by using the fact that the partition function of the model
is dominated by a single diagram of the high-temperature expansion, which is
calculated by an efficient combinatorial optimization algorithm. For a given
finite sample with discrete randomness the free energy is a piecewise linear function
of the temperature, which is rounded after averaging, however the discontinuity
of the internal energy at the transition point (i.e. the latent heat) stays finite
even in the thermodynamic limit. For a continuous disorder, instead, the latent heat
vanishes. At the phase transition point the dominant diagram percolates and
the total magnetic moment is related to the
size of the percolating cluster. Its fractal dimension
is found $d_f=(5+\sqrt{5})/4$ and it is independent of the type of the lattice and the form of
disorder. We argue that the critical behavior is exclusively determined by disorder
and the corresponding fixed point is the isotropic version of the so called infinite
randomness fixed point, which is realized in random quantum spin chains. From this
mapping we conjecture the values of the critical exponents as $\beta=2-d_f$,
$\beta_s=1/2$ and $\nu=1$.

\end{abstract}
\maketitle

\newcommand{\bc}{\begin{center}}
\newcommand{\ec}{\end{center}}
\newcommand{\be}{\begin{equation}}
\newcommand{\ee}{\end{equation}}
\newcommand{\beqn}{\begin{eqnarray}}
\newcommand{\eeqn}{\end{eqnarray}}

\section{Introduction}

Quenched disorder has generally a softening effect on the singularities of phase transitions
in homogeneous systems. As an example we mention that in the two-dimensional (2d) Ising model
the ordered phase (and thus the phase transition) is washed out by random fields\cite{aizenmanwehr,imrywortis} whereas due to
layered bond randomness the Onsager logarithmic singularity in the specific heat of the homogeneous
system will turn into a very weak essential singularity (McCoy-Wu model)\cite{mccoywu}. Singularities at first-order phase transitions are also soften\cite{Cardy99}. In 3d the latent heat is either reduced by a finite extent (for weak
disorder) or the latent heat completely disappears and the transition becomes second order (for sufficiently
strong disorder)\cite{3d}. In 2d, according to the rigorous result
by Aizenman and Wehr\cite{aizenmanwehr} there is no phase coexistence in the presence of continuous disorder, thus the phase transition is always continuous.  Thus there exists a set of random fixed points whose possible
classification is a great challenge of statistical physics. However, as in the theory of
disordered systems, exact results in this field are scarce and therefore our present
understanding of the subject is limited. Since a weak-disorder perturbation calculation starting
from the pure system's fixed point is not feasible most of the existing results are numerical. In
this respect an important role is played by the $q$-state ferromagnetic Potts model\cite{Wu}, in which the homogeneous
phase transition is of first order for $q>4$ in 2d\cite{baxter} and for $q \ge 3$ in 3d.

According to Monte Carlo\cite{pottsmc} and transfer matrix\cite{pottstm} investigations the critical behavior of the 2d random bond Potts model (RBPM) is governed by a line of $q$-dependent fixed points. For a given
$q$ the critical exponents are expected to be disorder independent, however generally there is a
strong cross-over regime\cite{earlymc}. The anomalous dimension of the magnetization, $x$, shows a smooth,
monotonously increasing $q$-dependence, which saturates at a given finite value as $q \to \infty$.
On the other hand the critical exponent of the correlation length, $\nu$, which is related to
energy-density correlations, is found to show only a very weak $q$-dependence. The measured
values are all around the rigorous bound\cite{ccfs}, $\nu=2/d=1$, which is required by general theorem
for disordered systems. (Note however the conflicting result in Ref.~\onlinecite{pottstm}, which is probably due to
the use of the bimodal, i.e. discrete form of disorder.)

Despite of the intensive numerical studies performed so far several aspects of the phase
transition in the RBPM are not clarified yet. Here we mention that it is still little
known about the mechanism which leads to the softening of the first order transition. The
effect of different types of disorder, continuous or discrete, has not been investigated.
Also universality of the transition with respect to the lattice structure is still an
open question. Finally, and most importantly, there is no example for a specific system
in which the critical properties are (conjecturedly) exactly known.

In this paper we are going to investigate the aforementioned questions using
the example of the 2d RBPM in the large-$q$ limit. The use
of this model has several advantages. i) The homogeneous model, which has a strongly
first-order phase transition, can be solved analytically\cite{Wu}. ii) In the presence of
randomness the model shows the generic features of disorder-rounded first-order phase
transitions. iii) It allows for an important technical simplification since the high-temperature
expansion of the model\cite{kasteleyn} (with arbitrary ferromagnetic couplings) is dominated by a
single diagram, whose calculation is reduced to an optimization problem\cite{JRI01}.
iv) Finally, this optimization problem is polynomial and there exists a powerful
combinatorial optimization algorithm\cite{aips02} which works in strongly polynomial time,
i.e. the necessary computational power does not depend on the actual form of the disorder.

This model, for which we shall refer simply as random bond Potts
model in the following, has already
been considered in several previous work. Cardy and Jacobsen\cite{pottstm} has shown a mapping
with the random-field Ising model in the level of
interface Hamiltonians, which explains the vanishing of the latent heat at the
transition point. Numerical studies for large (but finite) $q$-values\cite{jacobsenpicco}
predicted a smooth and non-singular behavior of the critical exponents as $q
\to \infty$. Relation of the random bond Potts model with an optimization problem was introduced
and studied by the method of simulated annealing in Ref.~\onlinecite{JRI01}. The optimization
problem was mathematically analyzed in Ref.~\onlinecite{aips02}. Here its
relation with sub-modular function optimization was shown and an efficient
computational algorithm was developed. Some numerical results obtained by the
optimization method has been briefly announced in a Letter\cite{ai03}.

The structure of the paper is the following. The model, its random cluster
representation and the way how its thermodynamical and correlational
properties are related to the dominant diagram are given in Sec.~\ref{S:II}.
In Sec.~\ref{S:III} the mechanism of
breaking of phase coexistence is analyzed  and an estimate of the breaking-up
length is given through extreme value statistics. Thermal quantities (internal energy,
specific heat) are calculated in Sec.~\ref{S:IV}. In particular we study the effect of discrete
and continuous disorder as well as the location of the transition point for different
2d lattices. The magnetization properties of the model, which are related to the
percolative properties of the largest connected cluster of the dominant
diagram is studied in Sec.~\ref{S:V}. In Sec.~\ref{S:VI} we show arguments and
approximate mappings between our model and random quantum spin chains which
leads us to conjecture the exact values of the critical exponents. Our conclusions
are drawn in the final section while two simple examples are
given in the Appendix.

\section{Formalism: The large $Q$-state Potts model}
\label{S:II}

The $q$-state Potts model\cite{Wu} is defined by the Hamiltonian:
\begin{equation}
\mathcal{H}=-\sum_{\left\langle i,j\right\rangle }J_{ij}\delta(\sigma_{i},\sigma_{j})
\label{eq:hamilton}
\end{equation}
in terms of the Potts-spin variables, $\sigma_{i}=0,1,\cdots,q-1$, at site $i$. The summation runs over all edges of a lattice $\langle i,j\rangle \in E$, and in our study the couplings, $J_{ij}>0$, are i.i.d. random variables. We consider the system in the large-$q$ limit
where, being the entropy per site $s \sim \ln q$,
it is convenient to introduce the reduced temperature, $T^{'}=T \ln q=O(1)$ and its inverse as $\beta^{'}=1/T^{'}=\beta/\ln q$. In the following we use these reduced variables but omit in the notation the prime. In terms of these variables the partition function is given by:
\begin{equation}
Z\equiv\sum_{\left\{ \sigma\right\} }q^{-\beta\mathcal{H}\left(\left\{ \sigma\right\} \right)}
\label{eq:Z}
\end{equation}
which is convenient to express in the random cluster representation\cite{kasteleyn} as:
\begin{equation}
Z =\sum_{G\subseteq E}q^{c(G)}\prod_{ij\in G}\left[q^{\beta J_{ij}}-1\right]
\label{eq:kasfor}
\end{equation}
Here the sum runs over all subset of bonds, $G\subseteq E$ and $c(G)$ stands for the number of connected components of $G$.

In the large-$q$ limit, where $q^{\beta J_{ij}} \gg 1$, the partition function can be written as
\begin{equation}
Z=\sum_{G\subseteq E}q^{\phi(G)},\quad \phi(G)=c(G) + \beta\sum_{ij\in G} J_{ij}\label{eq:kasfor1}
\end{equation}
which is dominated by the largest term, $\phi^*=\max_G \phi(G)$, so that
\begin{equation}
Z= n_0 q^{\phi^*}(1+ \dots).\label{eq:opti}
\end{equation}
Here the degeneracy of the optimal set $G^*$ is $n_0=O(1)$ and the omitted corrections in the r.h.s. go to zero for large $q$. The free-energy per site is proportional to $\phi^*$ and given by:
\begin{equation}
-\beta f= \frac{\phi^*}{N}\label{eq:free-e}
\end{equation}
where $N$ stands for the number of sites of the lattice.
In the Appendix 
we give an illustration of the convergence with $q$
of the family of function $f_q(\beta)$.

All information about the random bond Potts model is contained in the optimal set $G^*$.
The thermal properties are calculated from $\phi^*$, whereas the magnetization and the correlation
function are obtained from its geometrical structure. First we note that correlation between two
individual sites is unity only if both
are in the same connected cluster, otherwise the correlation is zero.
In this way the {\it average} correlation function, $C(r)$, is related to the distribution
of clusters in the optimal set.
Next we note that
if, and only if, there is an infinite cluster the correlation function
approaches a finite asymptotic value, $\lim_{r \to \infty}C(r)=m^2$. The magnetization $m$,
defined in this way, is the fraction of sites in the infinite cluster.
In the ferromagnetic phase, $T<T_c$, the magnetization is
$m>0$ and vanishes at the critical point, $T_c$, as  $m \sim (-t)^{\beta}$, where the relative
distance from the phase transition point is measured by $t=(T-T_c)/T_c$. As usual the
divergence of the correlation length $\xi$ at the critical point is given in the form:
$\xi \sim |t|^{-\nu}$,
where $\xi$ is defined as the average size of the clusters.
At the critical point the largest cluster is a fractal, its mass $M$ scales with the linear
size of the system $L$ as $M \sim L^{d_f}$, where $d_f$ is the fractal dimension.
Spin-spin correlations at the critical point decay algebraically as $C(r) \sim r^{-2x}$,
where the magnetization scaling dimension is given by: $x=d-d_f=\beta/\nu$.

In a system with a free surface one can define surface quantities, whose singularities
are governed by surface exponents. In the ferromagnetic phase the fraction of surface
sites belonging to the largest cluster defines the surface magnetization, which scales
as: $m_s \sim (-t)^{\beta_s}$. At the critical point surface-surface correlations
decay as $C_s(r) \sim r^{-2x_s}$ and the surface scaling dimension $x_s$ is related
to the surface fractal dimension $d_f^s$ as $d-1=d_f^s+x_s$. We have also $\beta_s=x_s \nu$.

In the random bond Potts model in the large-$q$ limit the structure and the fractal properties of the
optimal set are dominated by disorder effects. Other examples of systems in which the disorder
is dominant at the critical point are random quantum spin chains\cite{fisher}, for which the
set of critical exponents is exactly known.
This analogy and a possible isomorphism between the fixed points of the two problems,
which will be discussed
in detail in Sec.~\ref{S:VI}, bring us to conjecture the value of the critical exponents for the random bond Potts model, which we list below.

The fractal dimension of the infinite cluster at the transition point and
the scaling dimension of the (bulk)
magnetization are related to the golden-mean ratio, $\phi=(1+\sqrt{5})/2$, as:
\begin{equation}
d_f=1+\phi/2,\quad x=1-\phi/2 \label{eq:d_f}
\end{equation}
The analogous surface quantities are conjectured to be:
\begin{equation}
d_f^s=1/2,\quad x_s=1/2 \label{eq:d_fs}
\end{equation}
Finally, the correlation length exponent is given by:
\begin{equation}
\nu=1 \label{eq:nu}
\end{equation}
thus saturating the rigorous bound for disordered systems:\cite{ccfs} $\nu \ge 1$.

We close this section by a few technical remarks.
The cost-function of the problem $\phi(G)$, given in Eq.~(\ref{eq:kasfor1}), is a supermodular
function, thus there exists a combinatorial optimization method\cite{gls81}
to maximize it.  For the particular expression of
$\phi(G)$ in the present model a specific efficient algorithm, the Optimal Cooperation
Algorithm,\cite{aips02} has been formulated. The largest random system we treated by this
method is a $512 \times 512$ square lattice.

The investigations in this paper are restricted to two dimensions.
In particular we work on the square lattice
for which the location of the critical point is known for some distribution of the couplings. For
$0 < \beta J_{ij}<1$ and for a symmetric distribution $P(\overline{J}+J)=P(\overline{J} - J)$,
where $\overline{J}$ is the mean coupling, the critical
point is given by self-duality as\cite{dom_kinz}:
\begin{equation}
 T_c=\frac{1}{\beta_c}=2 \overline{J} \label{eq:T_c}
\end{equation}
 In the numerical calculations we used either a discrete (bimodal) distribution:
\begin{equation}
P_b(J)= p\delta(\overline{J} -\Delta-J)+ (1-p) \delta(\overline{J} +\Delta-J) \label{eq:bimodal}
\end{equation}
which is symmetric for $p=1-p=1/2$, or a continuous (uniform) distribution:
\begin{equation}
P_u(J)= \cases{1/2\Delta, \quad {\rm for}~ \overline{J}- \Delta<J<\overline{J}+\Delta\cr
               0, \quad {\rm otherwise} \cr} \label{eq:uni}
\end{equation}
In practical applications the latter distribution is approximated by a set of discrete peaks
(see for example Eq.~(\ref{eq:m_modal})), which is necessary since the algorithm works in terms of
rational numbers.
In the numerical calculations we collected data about systems up to a linear size $L=256$, The
number of realizations were about $10^6$ for the smaller systems and around $10^3$ for the largest one.
To check universality, we considered non-self-dual randomness,
analyzing triangular and hexagonal lattices
with number of sites up to 16384.

\section{Destruction of phase coexistence due to disorder}
\label{S:III}

In this section we consider the phase transition in the non-random system and see how
introduction of disorder will destroy the phase coexistence, i.e. soften the first order phase
transition into a second-order one. In particular in 2d we will estimate the disorder
dependent length-scale $l_b$, at which size the phase coexistence no longer exists.

The solution of the optimization problem in Eq.~(\ref{eq:opti}) depends on the temperature:
the optimal set is generally more interconnected in low temperature and contains more
disconnected parts in higher temperature.
In the homogeneous, non-random model with $J_{ij}=J$ there are only two (trivial), homogeneous
optimal sets: the fully connected diagram and the empty diagram. Consequently the free
energy of the homogeneous system ($f_{\rm hom}$) is given by:
\be
-N\beta f_{\rm hom}=\cases{1+N \beta Jz, \quad T<T_c \cr
                    N, \quad T>T_c \cr}
\label{pure}
\ee
where $z$ is the connectivity of the lattice.
At the phase-transition point, $\beta_c=1/T_c=(1-1/N)/(Jz)$, there is phase coexistence,
thus the phase transition is of first order and the latent heat, $\Delta e$, has its possible maximal
value: $\beta_c \Delta e=1$. Note that this result holds for any regular (even finite) lattice.

As quenched disorder is switched on, say in the paramagnetic phase, its fluctuations will locally
prefer the existence of the non-stable fully connected homogeneous diagram, and this tendency is
stronger around the phase-transition point, where the free-energy difference between the two
homogeneous phases is small. This mechanism, as the temperature is lowered, will
result in a sequence of non-homogeneous optimal sets with increasing fraction
of connected component, whose
average size is just the correlation length $\xi$. If $\xi$ stays finite at the phase transition
point then there is phase-coexistence and thus the first-order nature of the transition persists.
This is the case in 3d for sufficiently weak disorder.

In 2d, however, for any small amount of continuous disorder the correlation length is divergent
at the transition point thus the phase transition soften to second order. This is true in the
thermodynamic limit, while in a finite system of linear size $L$ weak disorder fluctuations
could be not sufficient to break phase coexistence. The 
finite length-scale, $l_b=L$, at which the breaking of phase-coexistence takes place can be
estimated through extreme value statistics in the following way\cite{note}.

At the transition point of the homogeneous system, i.e. at $\beta=1/(Jz)$, we compare the stability
of the empty graph with a non-homogeneous diagram $G_S$, in which all sites are isolated except of a domain of $S$ spins, which are fully connected. The relative cost-function is given by:
\begin{eqnarray}
W &=&\beta \sum_{ij\in G_S} J_{ij}-S+1\nonumber\\
&=& \beta \sum_{ij\in G_S} (J_{ij}-J)+ \beta J(Sz-\alpha S^{1/2})-S+1\nonumber\\
&\simeq& \frac{S^{1/2}\Delta}{z J}(\zeta -J\alpha/\Delta)+1\label{eq:W}
\end{eqnarray}
where we used that the distribution is symmetric, with $\overline{J}=J$.
Here in the second line the number of occupied edges of $G_S$ is $Sz$ minus the missing bonds
at the perimeter of the fully connected domain, which is given by $\alpha S^{1/2}$. In the last
equation we used that we are at the critical temperature and introduced the representation
$\sum_{ij}(J_{ij}-J)=\Delta S^{1/2} \zeta$, in terms of the Gaussian random
variable $\zeta$, which has zero mean and variance one.

Next we consider all independent positions of the fully connected domain in the finite lattice,
whose number is denoted by $N_S$, and look for the largest value of the cost-function, $W_0$,
in these diagrams. According to extreme value statistics\cite{galambos} in the large $N_S$ limit
its value is related to the cumulative probability distribution, $P(W>W_0)$, and follows from the
equation:
\begin{equation}
N_S P(W>W_0)=1 \label{eq:extr}
\end{equation}
Making use of the relation for a Gaussian variable: $P(\zeta>\zeta_0) \simeq
\exp(-\zeta_0^2/2)/\zeta_0$, we obtain from Eq.~(\ref{eq:extr}) that $W_0>0$, i.e. the
new diagram has a larger cost-function, than the empty set, if $N_S > \exp({\rm const}
(J/\Delta)^2)$. Here we are interested in the case when $S$ is in the same order as $L^2$,
i.e. when there is a complete breaking of the homogeneous diagram, in which case $N_S$ scales
as $N_S \sim L^{a}$, with $0<a<2$. Thus the breaking-up length-scale is just $l_b \sim L$
and given by:
\begin{equation}
l_b \approx l_0 \exp\left[ A \left( \frac{J}{\Delta} \right)^2 \right]\;,
\label{eq:l_b}
\end{equation}
where $l_0$ is a reference length.

{
\begin{figure}[h]
\centerline{\epsfxsize=3.25in\ \epsfbox{
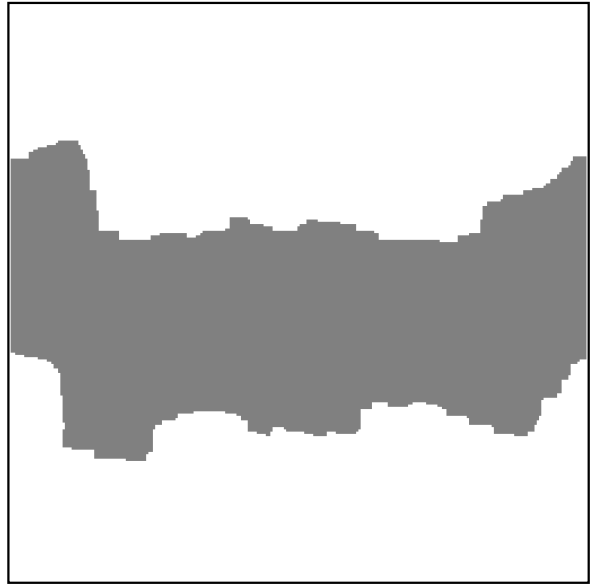}
}
\caption{Non-homogeneous optimal diagram for the bimodal distribution at a size L=256, 
corresponding to the breaking up disorder, $\Delta/J=693/2693=0.2573$. 
The connected part of the diagram is compact and percolates in one direction.
}
\label{Abb1}
\end{figure}
}

The relation between $l_b$ and $\Delta$ has been studied numerically, by calculating the
smallest $\Delta$ at
which the breaking of phase coexistence in a system of size $L$ takes place. 
(We used periodic boundary conditions.) 
An example of a non-homogeneous optimal diagram obtained 
in this way for the bimodal disorder is illustrated in Fig.~\ref{Abb1}.

{
\begin{figure}[h]
\centerline{\epsfxsize=3.25in\ \epsfbox{
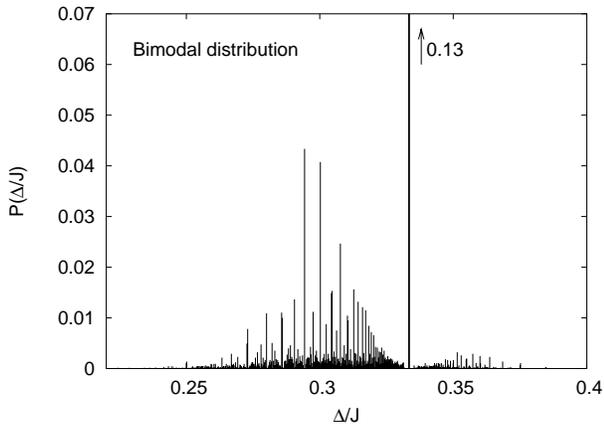}
}
\caption{
Distribution of the breaking up strength of bimodal disorder for a size $L=64$. The large individual peak at $\Delta/J=1/3$ corresponds to an instability due to a $3 \times 3$ plaquette.
}
\label{Abb2}
\end{figure}
}

For the bimodal disorder the distribution of the limiting values of $\Delta$ for a given size $L=64$
is shown in Fig.~\ref{Abb2}. The distribution consists of several individual peaks, some of those are related to instabilities due to small finite plaquettes. The size of the smallest such plaquette, $l$, is given by
$l \sim J/\Delta$, but the typical size of a connected domain varies between $l$ and $L$, thus there are strong sample to sample fluctuations. We have checked the validity of the relation in Eq.~(\ref{eq:l_b}) for the bimodal distribution and the result is plotted in Fig.~\ref{Abb3}. In the region of size we could work $L \le 128$ the
asymptotic behavior in Eq.~(\ref{eq:l_b}) has not yet been reached, the data points deviate from the expected straight line. The effective behavior of the breaking up length for these sizes is better described
(the relative error is 0.003) by a dependence $\ln L \sim J/\Delta$, as seen in the inset.
In the numerical calculation $J/\Delta$ goes to infinity faster than the prediction given above in Eq.~(\ref{eq:l_b}), 
because we implicitly consider in Eq.~(\ref{eq:W}) an euclidean domain $S$ whereas the Optimal Cooperation algorithm
finds more complicated domains.

{
\begin{figure}[h]
\centerline{\epsfxsize=3.25in\ \epsfbox{
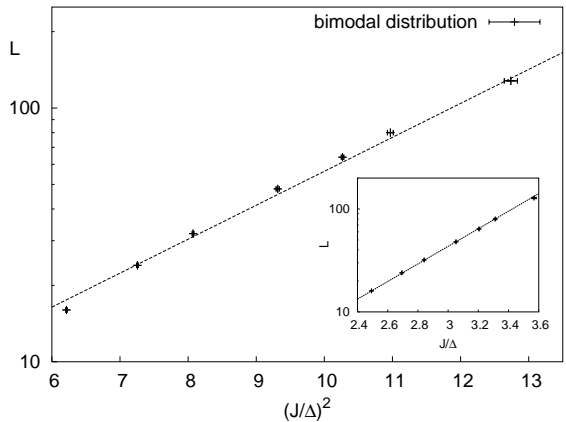}
}
\caption{
System size, $L$, as a function of the inverse average breaking-up disorder, $(J/\Delta)^2$, for bimodal distribution.
(In the average of $J/\Delta$ we considereded the contribution of all peaks). 
For the sizes we considered a fit with $(J/\Delta)$ is better, as shown in the inset.
}
\label{Abb3}
\end{figure}
}

We have repeated this calculation by using a (quasi) continuous disorder, which is approximated by
$127$ discrete
peaks. As seen in Fig.~\ref{Abb4} now the distribution of the breaking strength $\Delta$,
for a relatively small size,
$L=16$, is Gaussian-like, however for a sizes $L\geq32$, there is a secondary structure
reflecting the effect of discreteness in the representation of disorder.
As can be seen in Table~I, the measured breaking strengths are consistent with the result in Eq.(\ref{eq:l_b}),
unless for the two larger sizes, where the effect of discreteness of the distribution becomes significant.

{
\begin{figure}[h]
\centerline{\epsfxsize=3.25in\ \epsfbox{
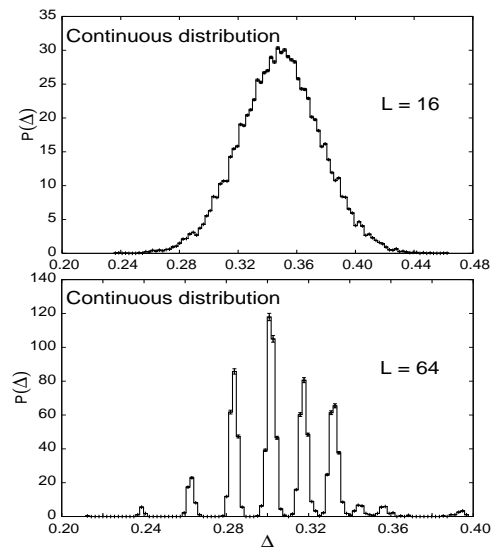}
}
\caption{
Distribution of the breaking up strength of quasi-continuous disorder composed of $m=127$
individual peaks
in Eq.~(\ref{eq:m_modal}) for two different sizes. For larger sizes the separated peaks are due to the
discreteness of the disorder.
}
\label{Abb4}
\end{figure}
}

\begin{table}
\begin{center}
\begin{tabular}{c c c c c }
\hline
$L$&  $<\frac{\Delta(L)}{J}>$& $\sigma_{\Delta/J}$ & $\left(\frac{\Delta}J\right)^2 \ln \frac L L_0$& samples\\
\hline
16 & 0.34854 &0.00016 & 0.3368 & 31324\\
24 & 0.3275  &0.0002  & 0.3408 &13927\\
32 & 0.3142  &0.0002  & 0.3422 &14786\\
48 & 0.3133  &0.0002  & 0.3801 & 15561\\
64 & 0.3075  &0.0004  & 0.3932 & 2878\\
\hline
\end{tabular}
\end{center}
\label{T:table}
\caption{Breaking strengths for the continuous disorder (for the fourth column $L_0 \simeq 1$ is used).}
\end{table}

In the numerical calculations on the critical properties of the RBPM, 
whose results are reported in the following sections, we always choose the disorder strong
enough, so that the relation $L \gg l_b$ is satisfied.

At this point we comment on a relation between the RBPM in the large-$q$ limit and the
random-field Ising model (RFIM) at $T=0$, which has already been observed by Cardy and
Jacobsen\cite{pottstm}. The form of the cost-function in Eq.~(\ref{eq:W}) is identical to
that of the RFIM, where bonds in the RBPM are equivalent to sites in the RFIM having a random
field of $J_{ij}-J$ and the perimeter contribution in Eq.(\ref{eq:W}) is related to the domain-wall
energy in the RFIM. Thus $\beta J$ corresponds to the ferromagnetic coupling in the RFIM.
This mapping on the level of the interface Hamiltonian is however restricted to a smooth,
non-fractal interface. Therefore it applies in the same way in the stability analysis
of the pure phases, resulting in a similar breaking-up length-scale as in Eq.~(\ref{eq:l_b}).
However the (fractal) structure of the clusters in the two problems are different.
As a consequence the bulk and surface fractal dimensions (and the related critical exponents)
are different in the two problems. For the RFIM the percolating clusters have the same
fractal properties as ordinary (site) percolation, as has been demonstrated by numerical
calculations\cite{sepp}.

\section{Singularities of thermal quantities for discrete and continuous disorder}
\label{S:IV}

In this section we analyze the singularities of thermal quantities, in particular we consider the
internal energy, $e$, and the specific heat, $c_V$. We start to rewrite Eq.~(\ref{eq:free-e}) for
the free energy of a given realization of disorder as:
\begin{equation}
F=-c(G^*)T - \sum_{ij\in G^*} J_{ij} \label{eq:free-e1}
\end{equation}
where the  optimal set, $G^*$, changes discontinuously as temperature is varied.
For sufficiently high temperature, $\beta{'}z ~{\rm max}(J)<1$, $G^*$ is the empty set, whereas for low temperatures,
$\beta{'}z ~{\rm min}(J)>1$, it is fully connected. In between, around the transition point,
however, there are more frequent changes. In a finite system of size $L$, the temperature
interval of stability of the optimal set is found numerically to be around $\sim 1/L$, for
the bimodal distribution. As a consequence in a finite system the free energy is a piecewise
linear function of the temperature, whose slope (the entropy, $c(G^*)$), is monotonously increasing with the temperature. From Eq.~(\ref{eq:free-e1}) the internal energy is obtained by
derivation, which for a given sample is:
\begin{equation}
E=- \sum_{ij\in G^*} J_{ij} \label{eq:ener}
\end{equation}
This is a piecewise constant function of the temperature, thus in a given sample, even for a
finite system the internal energy shows discontinuities. We illustrate this point in a simple example in the Appendix, where
 the comparison between the bimodal and
the uniform distribution for a general simple triangle is given.
In the given (finite) sample the phase
transition point separates the two qualitatively different regimes in which the largest cluster
is not percolating (paramagnetic phase), from that in which it is percolating (ferromagnetic phase).
This phase transition in the given sample is sharp and of first order. Later we shall use this
definition to locate the transition point of a series of samples and study their distribution.

In the random system, as we mentioned before, the average quantities, such as the average
free energy and its derivatives are of physical relevance. The averaging procedure
generally acts to smear out the discontinuities
in the internal energy. In this respect the behavior of the averaged quantities is
different for the discrete and the continuous distributions. (See also the example in the Appendix.) As illustrated in Fig.~\ref{Abb5}
for the bimodal distribution several discontinuities remain in the average internal energy,
whereas for the uniform distribution the average internal energy is continuous,
as illustrated in Fig.~\ref{Abb6}. For the discrete distribution the origin of the discontinuities
is the degeneracies of $\phi$ at some temperature.
For example
consider a square lattice where the bonds follow the bimodal distribution
in Eq.~(\ref{eq:bimodal}). In this lattice let us focus on an elementary square, whose
bonds are all strong. The sites of this plaquette are either non connected
(giving a contribution to $f$ of $4 T$) or are fully connected (with a contribution to $f$ of
 $T+4(\overline{J}+\Delta)$). These two situations are degenerate at $T=4/3 (\overline{J}+\Delta)$.
The average jump in the internal energy, $\Delta e$,
at this temperature can be estimated by the approximation with independent plaquettes,
where a fraction of $1/8$ squares have the above degeneracy, each of which giving a local
jump per site of $3 T/4$. Thus $\Delta e \approx 3 T/32$ which agrees reasonably well with
the measured jump in Fig.~\ref{Abb5}.

\begin{figure}
  \begin{center}
     \includegraphics[width=2.35in,angle=270]{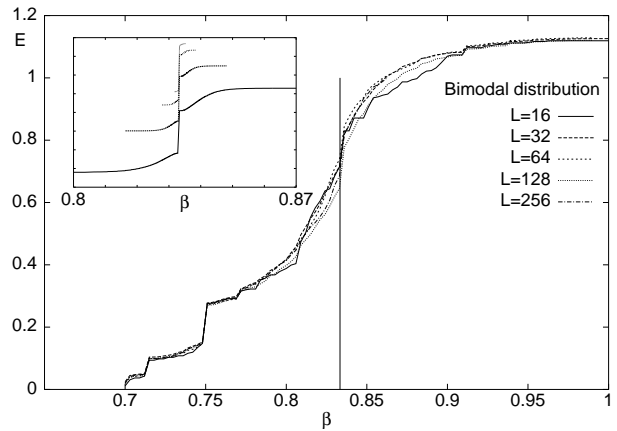}
   \end{center}
   \caption{Internal energy with bimodal disorder ($\beta J_1 = 1/6$ and
	$\beta J_2 = 5/6$) as a function of the
inverse temperature, $\beta$. Note on the discontinuities which are present after averaging in the
thermodynamic limit, c.f. at $\beta=0.75$. In the inset the discontinuous behavior at the transition
point, $\beta=5/6$, is illustrated (the lines from bottom to top are for increasing size from $L=32$ to
$L=256$).}
   \label{Abb5}
 \end{figure}

The average internal energy displays also a finite jump at the transition point,
as illustrated in the inset to Fig.~\ref{Abb5}. For the bimodal distribution this jump comes
from those configurations in which a corner site has a strong and a weak bond, so that
the isolated corner and the connected corner has the same contribution. The fraction of
these corners is finite at the transition point, which leads to a finite average jump.
Note, however, that the positions of these degenerate corners are distributed overall in the sample,
thus their contribution is not exclusively related to the appearance of a percolating cluster.

{
\begin{figure}[h]
\centerline{\epsfxsize=3.25in\ \epsfbox{
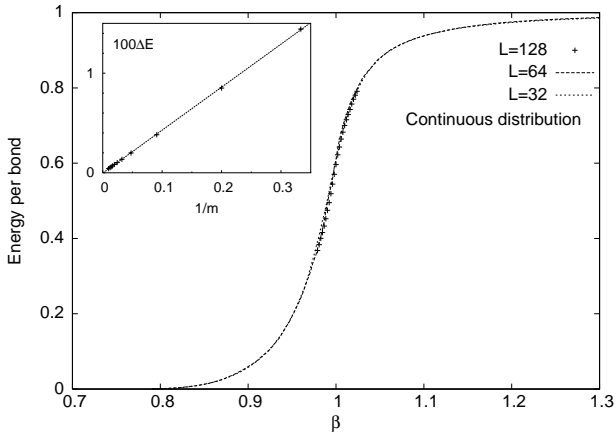}
}
\caption{
Internal energy for (quasi-)continuous disorder as given in Eq.~(\ref{eq:m_modal}). In the inset
the latent heat is plotted as a function of the inverse number of the discrete peaks in the distribution.
}
\label{Abb6}
\end{figure}
}

The jumps in the average internal energy will disappear, if a continuous distribution of disorder is used. To illustrate it we used a distribution, which consists of $2m$ discrete peaks of equal weight:
\begin{eqnarray}
P_m(J)&=& \frac{1}{2m} \sum_{k=0}^{m-1}\left[ \delta\left(\overline{J}
-\Delta-\left((m-1)/2 - k\right)\varepsilon-J\right)\right.\nonumber\\
&+&\left. \delta\left(\overline{J} +\Delta+((m-1)/2-k)\varepsilon-J\right)\right] \label{eq:m_modal}
\end{eqnarray}
so that in the large-$m$ limit we approach a continuous distribution. In Eq.~(\ref{eq:m_modal})
$\varepsilon$ is a small number, and in the simulation it has been used
$\varepsilon = 2 \overline J 10^{-5}$.
As shown in the inset to
Fig.~\ref{Abb6} the jump in the average internal energy at the transition point goes to zero
as $\sim 1/m$, which can be understood as follows. For a given corner site out of $m^2$ different
 coupling sets there are only $m$ for which the degeneracy, as described for the bimodal
distribution holds. Therefore the average jump should scale with $1/m$, as observed.

Going back to the internal energy for the bimodal distribution we argue that the true singularity can be observed at two sides of the transition point, i.e. as $t \to 0^+$ or $t \to 0^-$. In a finite system of length $L$, we expect a singularity in the form:
\begin{equation}
e(t,L)=e(0^+,L)+t^{1-\alpha} \tilde{e}(tL^{1/\nu}),\quad t > 0 \label{eq:e_tL}
\end{equation}
and similarly for $t \le 0$. Here the specific heat exponent, $\alpha$, satisfies the hyperscaling relation: $d \nu = 2 - \alpha$. The scaling function $\tilde{e}(y)$, is expected to be an odd function, thus, for small argument, to behave as $\tilde{e}(y) \sim y$. The scaling prediction in Eq.~(\ref{eq:e_tL}) has been checked and the results are shown in Fig.~\ref{Abb7}.
The scaling collapse of the data is indeed very good with the exponents: $\nu=1$ and $\alpha=0$, as conjectured in Eq.~(\ref{eq:nu}). From the collapse of the points we estimate $\nu=1$ within a numerical precision of $1\%$.

{
\begin{figure}[h]
\centerline{\epsfxsize=3.25in\ \epsfbox{
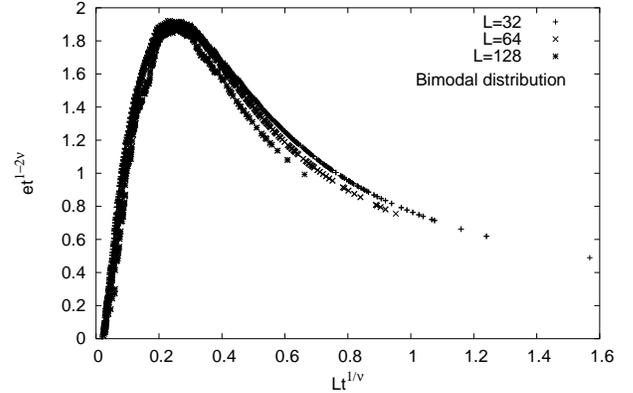}
}
\caption{
Scaling plot of the singular part of the internal energy in Eq.~(\ref{eq:e_tL}) for bimodal disorder. In the plot there is one free parameter, $\nu$, for which we used the conjectured value in Eq.~(\ref{eq:nu}).}
\label{Abb7}
\end{figure}
}

As we mentioned below Eq.~(\ref{eq:ener}) one can define a transition temperature, $T_c(L)$,
in a given finite sample, which separates the regime in which there is a percolating cluster
from that in which the largest cluster is not percolating. 
(We note that the percolating cluster at strong enough disorder, i.e. when
$l_b \ll L$, is fractal, in contrary to the cluster at the breaking-up
strength of disorder in Fig.~\ref{Abb1}, which is compact.)
The distribution of this temperature
for bimodal and continuous disorder is shown in Fig.~\ref{Abb8} for periodic boundary conditions. 
For large systems one expects
that the distributions approach a unique limiting curve in terms of the scaled variable,
$\tau=t_c /\sigma$, where $t_c=T_c(L)-T_c$ is the distance from the critical temperature
in the infinite system and the variance of the distribution scales as $\sigma(L) \sim L^{-\omega}$,
with some scaling exponent, $\omega$. As seen in Fig.~\ref{Abb8} this limiting distribution, $P(\tau)$,
has its maximum in $\tau>0$ and the integrated distribution is :
$P_{\rm sp}=\int_0^{-\infty} P(\tau){\rm d} \tau>1/2$. As a matter of fact
$P_{\rm sp}$ is the spanning probability given by the fraction of samples in which the
largest cluster is percolating.
Indeed a large (but finite) sample percolating at $\tau>0$, is also percolating at $T=T_c$.
According to our numerical results for the bimodal distribution $P_{\rm sp}=0.63(1)$.
The finite size corrections to the spanning probability are given by: $P_{\rm sp}-P_{\rm sp}(L) \sim
\Delta t(L)$, where $\Delta t(L) \sim L^{-\epsilon}$ is the shift of the critical temperature.
Here we note that in a pure system the shift exponent is generally given by: $\epsilon=1/\nu$,
if the transition is of second order. If, however the transition is of first-order the critical
exponent is
given by its discontinuity fixed point value\cite{disc_fp,turbanigloi02}, which is $\nu_{d}=1/d$.
This is in accordance with the exact result in the pure model, in which the shift in the transition
temperature is given by $\sim 1/N$, as given below Eq.~(\ref{pure}).
From the available data
we could not make an independent estimate for $\epsilon$, however the data points seem to
be consistent with $\epsilon=1/\nu=1$
for bimodal randomness. The exponent $\omega$ could be determined also with relatively large
errors giving an estimate $\omega=0.9(1)$.

{
\begin{figure}[h]
\centerline{\epsfxsize=3.25in\ \epsfbox{
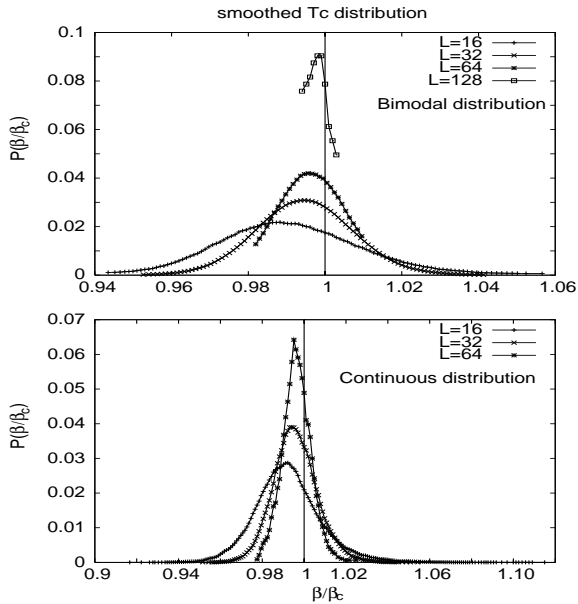}
}
\caption{
Distribution of the finite-size (percolation) transition temperature for the bimodal (a) and the continuous (b) disorder for different sizes of the system. Note that the difference from the exact transition
temperature is in the same order as the variance.
}
\label{Abb8}
\end{figure}
}

\begin{figure}
  \begin{center}
     \includegraphics[width=3.25in]{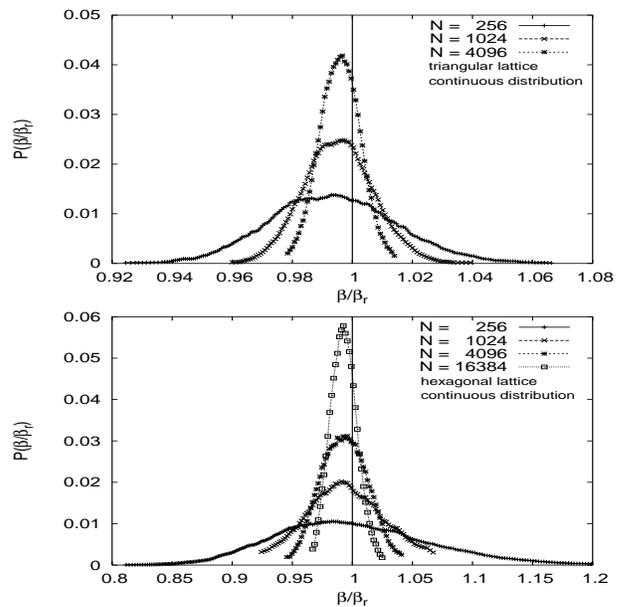}
   \end{center}
   \caption{Distribution of the finite size transition (percolation) temperature for triangular and hexagonal lattices 
with continuous distribution of disorder for different sizes.}
   \label{AbbHex}
 \end{figure}

We have also studied the location of the critical point on the triangular lattice,
for which the coordination number is $z=3$ and on the hexagonal lattice, which has
$z=3/2$. In Fig.~\ref{AbbHex} we present the distribution of the finite-size critical
temperatures for different values of the number of sites $N$ for the triangular and hexagonal lattice,
with symmetric and continuous disorder and for periodic boundary conditions. 
Here we take as reference temperature
$\beta_r=1/(z \overline{J})$, which is the transition point for the non-random
system and expected to be a good approximation for symmetric disorder.
The spanning probability at the reference temperature is $P_{sp}=0.65(5)$ for both kind of lattices,
which is close to that in the square lattice.

\section{Magnetization and the fractal properties of the percolating cluster}
\label{S:V}

Order and correlations in the RBPM are related to the structure of clusters in the optimal set $G^*$.
As we explained in Sec.~\ref{S:II} at the critical point the largest cluster is a fractal,
its mass $M$ scales with the linear size of the system $L$ as $M \sim L^{d_f}$. The structure of a
typical optimal set is illustrated in Fig.~1 of Ref.~\onlinecite{ai03}  for the bimodal distribution.
Note that in $G^*$ each connected component contains all the possible edges,
since the inclusion of any coupling with $J_{ij}>0$ will increase the cost function.
The topology of clusters in ordinary bond percolating is evidently different, since
they contain all the strong bonds but none of the weak ones.

\begin{figure}
  \begin{center}
     \includegraphics[width=2.35in,angle=270]{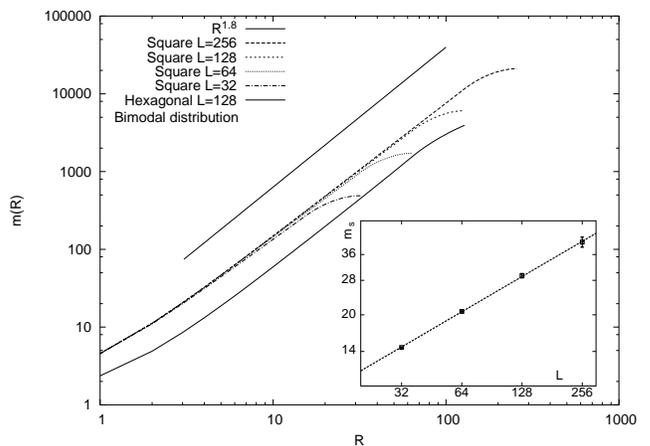}
   \end{center}
   \caption{
Average mass of the percolating cluster within a region of size $r$ at the critical point, for the square and the hexagonal lattices. The conjectured asymptotic behavior is indicated by a straigth line. Inset: average mass of surface points of the percolating cluster in the square lattice.
}
   \label{Abb9}
 \end{figure}

We checked numerically that the largest cluster is indeed a fractal by measuring the number of points
in the cluster, $\mu(r,L)$, which are at most at distance $r$ from a reference point. Similarly we
measured the number of points in the cluster between a distance $r+1$ and $r$: $s(r,L)\equiv
\mu(r+1,L)-\mu(r,L)$. Here we performed averaging i) over the position of the reference point and ii)
over the disorder realisations. Under a scaling transformation, $r \to r/b$, $\mu(r,L)$ and $s(r,L)$
are expected to behave as:
\begin{eqnarray}
 \mu(r,L)=b^{d_f} \mu(r/b,L/b) \nonumber\\
 s(r,L)=b^{d_f-1} s(r/b,L/b)
 \label{eq:mu_s}
\end{eqnarray}
Now in the first equation taking $b=r$ we obtain, $\mu(r,L)=r^{d_f} \tilde{\mu}(L/r)$, thus in a
log-log plot one should obtain a straight line for $L > r$. As seen in Fig.~\ref{Abb9}
this relation is indeed
satisfied both for the square and for the hexagonal lattices, in both cases - within the error of
 the calculation - having the same asymptotic slopes, which are compatible with the conjectured result
in Eq.~(\ref{eq:d_f}).
We also performed a multifractal analysis showing only a single fractal dimension.

\begin{figure}
  \begin{center}
     \includegraphics[width=2.35in,angle=270]{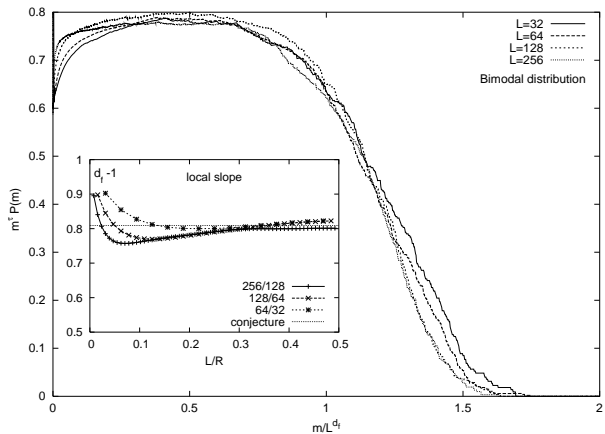}
   \end{center}
   \caption{
Scaling plot of the cumulative distribution of the mass of the clusters at the critical point, using the conjectured value of the fractal dimension in Eq.~(\ref{eq:d_f}). In the inset the fractal dimension is presented as calculated from the ratio of the masses of the percolating clusters in different finite systems (see text).
}
   \label{Abb10}
 \end{figure}

To have a more precise estimate of the fractal dimension we calculated $s(r,L)$ for finite systems of size $L=16,32,64,128$ and $256$. Comparing results of two consecutive sites we obtained $d_f$ from the second equation of Eq.~(\ref{eq:mu_s}) by setting $b=2$. The effective fractal dimensions as plotted in the inset to Fig.~\ref{Abb10}  have only a weak $r/L$ and $L$ dependence, and from the asymptotic behavior one can estimate $d_f=1.81(1)$ in good agreement with the conjectured result in Eq.~(\ref{eq:d_f}).

Next we studied the cumulative distribution of the mass of the clusters, $R(M,L)$, which measures
the fraction of clusters having at least a mass of $M$. According to scaling
theory\cite{staufferaharony} it should asymptotically behave as:
\begin{equation}
R(M,L)=M^{-\tau} \tilde{R}(M/L^{d_f})\label{eq:R_mL}
\end{equation}
with $\tau=(2-d_f)/d_f$. Note that the scaling relation in Eq.~(\ref{eq:R_mL}) contains only one
free parameter, $d_f$, which can be estimated from an optimal scaling collapse of the distributions
calculated for different sizes. As shown in Fig.~\ref{Abb10}  the scaling collapse is excellent if the conjectured
fractal dimension in Eq.~(\ref{eq:d_f}) is used. By analyzing the accuracy of the collapse we obtained
an estimate: $d_f=1.810(5)$ in very good agreement with the conjectured value.

We close this section by studying the surface magnetization of the model. Using open boundary conditions we measure the average mass of the surface sites belonging to the largest cluster, $M_s$. According to considerations in Sec.~\ref{S:II}  in a finite system at the critical point this should asymptotically behave as: $M_s \sim L^{d_f^s}$. According to the data presented in the inset to Fig.~\ref{Abb9} the surface fractal dimension is given by
$d_f^s=0.495(10)$, which is in good agreement with the conjectured value in Eq.~(\ref{eq:d_fs}).

\section{Anisotropic randomness and relation with the strong disorder
fixed point of random quantum chains}
\label{S:VI}

In this section the random bond Potts model is related to another problem of statistical
physics, which has partially exact results about its critical properties. The analogies and
relations between the models will be used to conjecture the values of the critical
exponents in the random bond Potts model,
which are already announced in Eqs.(\ref{eq:d_f}),(\ref{eq:d_fs}) and (\ref{eq:nu}).

The problem we consider first is the Potts model with correlated bond disorder, in which the vertical bonds are
constant denoted by $J_{\parallel}$, whereas the horizontal couplings have the same
value in a given column, $i=1,2,\dots,L$, denoted by $J_i$. Here the $J_i$ are i.i.d. random variables and in the following we choose $J_{\parallel}=\overline{J}$. This model, which for Ising spins is the well known McCoy-Wu model\cite{mccoywu}, is translationally symmetric in the vertical direction. Consequently
the optimal set has a strip-like structure, and the 2d diagram $G$, is uniquely characterized by its 1d cut,
denoted by $g$. The cost-function of the problem can be written as $\phi(G)=L_{\parallel} \psi(g)$,
where the size of the system in the vertical (horizontal) direction is $L_{\parallel} \sim L$ ($L$) and
\be
\psi(g)=c_1(g) + \frac{1}{L_{\parallel}}c_2(g) + \beta \sum_{ij\in g} J_{ij}
+\beta \overline{J}(L-c_1(g))\;,
\label{psi}
\ee
is the cost-function of the 1d problem. Here $c_1(g)$ is the number of isolated points and $c_2(g)$ is the number
of connected diagrams which has at least two sites. In the thermodynamic limit the second term in the r.h.s is
negligible and redefining the temperature we arrive to an equivalent cost-function:
\be
\tilde{\psi}(g)=c_1(g) +  \tilde{\beta}\sum_{ij\in g} J_{ij}\;,
\label{psi1}
\ee
with $\tilde{\beta}=\beta
/(1-\beta \overline{J})$ and at the critical point  $\tilde{\beta}\overline{J}=1$.The optimization problem in Eq.~(\ref{psi1}) will be the subject of a separate publication.

Next we consider the extreme anisotropic\cite{kogut}
(time-continuum or Hamiltonian) limit of the strip-random Potts model,
where we let
$J_{\parallel}/\overline{J}\to 0$
and the transfer matrix of the model in the vertical direction is written as:
$\cal{T}=\exp(-\tau \cal{H})$. Here $\tau$ is the (infinitesimal) lattice spacing and $\cal{H}$ is the Hamiltonian
of a random Potts chain in the presence of a transverse field\cite{turbanigloi02}:
\be
{\cal H}=-\sum_i J_i \delta(s_i,s_{i+1})-\sum_i \frac{h}{q} \sum_{k=1}^{q-1} M_i^k\;,
\label{Hpotts}
\ee
where $M_i | s_i \rangle=| s_i+1, {\rm mod}~q \rangle$. If the couplings in vertical lines vary randomly from line to line then the transverse fields, $h_i$, are random variables, too. The fixed point of the 2d strip-random Potts model and that of the random quantum Potts chain with Hamiltonian in Eq.~(\ref{Hpotts}) are isomorph, thus they have the same set of critical exponents. The latter system has been
studied in detail by a strong disorder renormalization group method, which was originally introduced by Ma,
Dasgupta and Hu\cite{mdh} and used later by Fisher\cite{fisher} and others\cite{senthil,ijl01,2drg}. This method is assumed to give asymptotically
exact results, in particular the critical exponents for $q \ge 2$ are identical\cite{senthil} and are given
by\cite{fisher}:
\be
\nu_{\perp}=2, \quad x_{\perp}=1-\phi/2, \quad x_{\perp}^s=1/2\;,
\label{QPexp}
\ee
which also hold for the 2d strip-random model.

Having the critical behavior in the presence of correlated, anisotropic randomness we try to relate it to
the original problem, in which the disorder is isotropic. To do so we imagine that at the fixed point
of the anisotropic problem we let the couplings to be random also in the vertical direction. In this way
translational symmetry in the vertical direction is broken and in the originally homogeneous strips connected
and disconnected parts will appear. At the critical point these two (creation and destruction) processes are
symmetric, therefore it is plausible to assume that the mass of the largest cluster stays invariant, thus:
$M \sim L_{\parallel}\times L^{1-x_{\perp}} \sim L^{d_f}$, from which the value of the fractal dimension,
$d_f=2-x_{\perp}$, as announced in Eq.~(\ref{eq:d_f}) follows. Repeating this argument for a surface cluster
we obtain the result in Eq.~(\ref{eq:d_fs}). Another, and related assumption is that the correlation volume, i.e. the surface of the largest cluster grows in the same way in the two problems, as the transition point is approached. This means, that $V(t) \sim \xi_{\parallel}(t) \xi_{\perp}(t) \sim L_{\parallel}
|t|^{-\nu_{\perp}} \sim \xi(t)^2 \sim |t|^{-2\nu}$, from which $\nu=\nu_{\perp}/2=1$ follows, in accordance
with the announced result in Eq.~(\ref{eq:nu}).

\section{Conclusion}

One of the most spectacular effect of quenched disorder is the rounding of first order transitions.
In 2d this phenomenon can not be studied in a perturbative basis around the pure systems fixed point.
In this paper we have demonstrated that an almost complete understanding of this phenomenon can be
achieved in the other limiting case, when the effect of disorder is strong and dominates the critical
behavior. We argued that the appropriate system for this purpose is the random bond ferromagnetic
Potts model in the large-$q$ limit, in which thermal fluctuations are strongly suppressed. In the
random cluster representation the properties of the system are related to the dominant diagram of
the high-temperature expansion. At the critical point this diagram is a self-similar fractal, giving
a natural explanation of the diverging length-scale, and its bulk and surface fractal dimensions are
related to the critical exponents of the transition. We have introduced a plausible mapping between
our isotropic system and that of strictly correlated disorder, which is isomorph with random quantum
spin chains for which the critical properies are known.
The critical properies  conjectured in this way  are then checked by large scale numerical
calculations based on a very efficient combinatorial optimization algorithm. We have studied in detail
the possible differences in the results obtained for discrete or continuous disorder.
For discrete randomness, such as the bimodal one, which is frequently used in numerical calculations the internal energy displays discontinuities, also the latent heat is finite. This observation
should make people cautious while analyzing data obtained by the use of discrete disorder.

The results of this paper can be extended in several directions. Including in the distribution of the couplings the values $\beta J_{ij}<0$ and  $\beta J_{ij}>1$ we obtain a system in which percolation and random bond effects compete. This problem can also be studied by our optimization algorithm and the properties of the stable fixed point can be calculated by analyzing the fractal properties of the relevant diagrams.
Another extension is to include negative couplings, thus a Potts spin-glass problem is obtained.
This still can be formulated as an optimization problem, which is most probably np-complete. Considering the ferromagnetic case one can try to perform a large-$q$ expansion around the conjecturedly exact results. For the critical exponents the corrections are expected to go as $1/\ln q$\cite{jacobsenpicco}.
Finally a further possible study is to clarify 
the dynamical properties of the system and to show if there are some relations
with the ultraslow dynamics present in random quantum spin chains\cite{fisher,bigpaper}.

We are indebted to R. Juh\'asz,
M. Preissmann, H. Rieger, A. Seb\H o and L. Turban for stimulating discussions.
This work has been supported by the French-Hungarian cooperation programme Balaton (Minist\'ere des
Affaires Etrang\`eres - OM), the Hungarian National
Research Fund under  grant No OTKA TO34183, TO37323,
MO28418 and M36803, by the Ministry of Education under grant No FKFP 87/2001,
by the EC Centre of Excellence (No. ICA1-CT-2000-70029).

\appendix
\section{}
\label{AppendaA}

Here we present two examples to illustrate the discontinuous nature of the internal energy in the $q \to \infty$ limit.

\begin{figure}[h]
\centerline{\epsfxsize=3.25in\ \epsfbox{
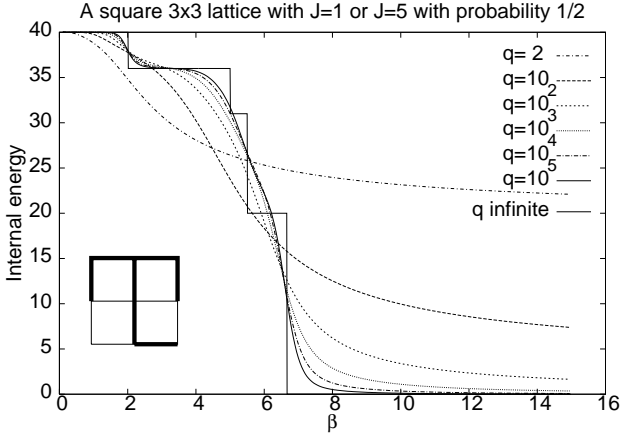}}
\caption{
Average internal energy as a function of $\beta$ for various values
of $q$. In the inset the analyzed plaquette is shown, where the thick lines represent strong bonds ($J=5$)
and the thin lines weak ones ($J=1$).}
\label{figapp}
\end{figure}

We start with a $3 \times 3$ plaquette of the square lattice in which the couplings are $1$ or $5$ with the same probability.
The realization we investigate is shown in the inset of Fig.~\ref{figapp}.
To illustrate the approach to $q \to \infty$ we have calculated for different values of $q$ the internal energy as a function of the reduced temperature ($T \to T \ln q$), as introduced below Eq.(\ref{eq:hamilton}). As seen in Fig.~\ref{figapp} for any finite $q$ the internal energy is continuous and becomes discontinuous only in the $q \to \infty$ limit.

Our second example is to illustrate the effect of rounding due to averaging over disorder. We consider here an elementary triangle in which the edges carry the (positive) weights: $a$, $b$ and $c$ ($a=\beta J_a$, etc.). By simple inspection one finds that,
for a given triplet $\left\{ a,b,c\right\} $, the free energy is:

\[
-\beta F_{a,b,c}(\beta )=
\left\{ \begin{array}{cc}
  \begin{array}{ccc}
 1+\beta (\widetilde{a}+\widetilde{b}+\widetilde{c}) & \textrm{if} & \widetilde{a}+\widetilde{b}+\widetilde{c}\leq \frac{2}{\beta }\\
 3 & \textrm{if} & \frac{2}{\beta }\leq \widetilde{a}+\widetilde{b}+\widetilde{c}\end{array}
  & \\ \end{array}
\right.
\]
when $\widetilde{a}+\widetilde{b}\leq \widetilde{c}$ and
\[
-\beta F_{a,b,c}(\beta )=
\left\{ \begin{array}{cc}

  \begin{array}{ccc}
 1+\beta (\widetilde{a}+\widetilde{b}+\widetilde{c}) & \textrm{if} & \widetilde{c}\leq \frac{1}{\beta }\\
 2+\beta \widetilde{c} & \textrm{if} & \widetilde{a}+\widetilde{b}\leq \frac{1}{\beta }\\
 3 & \textrm{if} & \frac{1}{\beta }\leq \widetilde{a}+\widetilde{b}\end{array}
\leq \widetilde{c} & \end{array}
\right.
\]
when $\widetilde{c}\leq \widetilde{a}+\widetilde{b}$.
We use the notation $\left\{ \widetilde{a},\widetilde{b},\widetilde{c}\right\} =\left\{ a,b,c\right\} $
and $\widetilde{a}\leq \widetilde{b}\leq \widetilde{c}$ (i.e. the
variables with the tilde are ordered in ascending order).

For the uniform distribution of disorder in Eq.(\ref{eq:uni}) with $\overline{J}=\Delta=1/2$ one finds for the {\it average} internal energy \[
E(T)=\left\{ \begin{array}{ccc}
 -\frac{1}{8}T^{4}-T^{3}+\frac{3}{2} &  & T<1\\
 -2T^{4}+8T^{3}-9T^{2}+\frac{27}{8} &  & 1<T<\frac{3}{2}\\
 0 &  & \frac{3}{2}<T\end{array}
\right.\]
which is a continuous function of the temperature.

By contrast for the discrete bimodal dsitribution in Eq.(\ref{eq:bimodal}) with $\overline{J}=1/2$ and $\Delta=1/\sqrt{3}$, which has the same mean and variance as the uniform distribution the internal energy is given by:\[
E(T)=\left\{ \begin{array}{ccc}
 0 &  & T<\frac{3}{4}-\frac{\sqrt{3}}{4}\\
 \frac{3}{16}-\frac{\sqrt{3}}{16} &  & \frac{3}{4}-\frac{\sqrt{3}}{4}<T<1-\frac{\sqrt{3}}{3}\\
 \frac{3}{2} &  & 1-\frac{\sqrt{3}}{3}<T<\frac{1}{2}+\frac{\sqrt{3}}{6}\\
 \frac{3}{4}-\frac{\sqrt{3}}{8} &  & \frac{1}{2}+\frac{\sqrt{3}}{6}<T<\frac{3}{4}+\frac{\sqrt{3}}{12}\\
 \frac{21}{16}-\frac{\sqrt{3}}{16} &  & \frac{3}{4}+\frac{\sqrt{3}}{12}<T<\frac{3}{4}+\frac{\sqrt{3}}{12}\\
 \frac{3}{2} &  & \frac{3}{4}+\frac{\sqrt{3}}{12}<T\end{array}
\right.\]
This presents five discontinuities of width 0.079, 0.296, 0.158, 0.671,
and 0.296 at the temperatures 0.317, 0.423, 0.789, 0.894, and 1.183, respectively.

\end{document}